\documentclass[aps,prl,twocolumn,showpacs,superscriptaddress,groupedaddress]{revtex4-1}
\DeclareMathAlphabet{\mathpzc}{OT1}{pzc}{m}{it}
\usepackage{amsfonts}
\def\Q{\textbf{Q}}
\def\X{\textbf{X}}
\def\g{\hat{\gamma}}
\def\ggg{\gamma}

\usepackage{mathrsfs}
\usepackage{graphicx}
\usepackage{amsmath}
\usepackage{epstopdf}
\usepackage{color}
\usepackage{soul}

\definecolor{darkblue}{RGB}{83,0,93}
\newcommand{\q}[1]{\textcolor{darkblue}{#1}}

\newsavebox{\astrutbox}
\sbox{\astrutbox}{\rule[-5pt]{0pt}{20pt}}

\newcommand\p{\ensuremath{\partial}}

\def\p{\partial}
\def\be{\begin{eqnarray}}
\def\ee{\end{eqnarray}}
\def\bes{\begin{subeqnarray}}
\def\ees{\end{subeqnarray}}

\def\f{\frac}
\def\lp{\left(}
\def\rp{\right)}
\def\lb{\left[}
\def\rb{\right]}

\def\n{\nabla}

\def\Q{\textbf{Q}}

\def\befi{\begin{figure}}
\def\eefi{\end{figure}}

\def\bce{\begin{center}}
\def\ece{\end{center}}

\def\h{\hat{}}
\def\RA{\rightarrow}
\def\d{\delta}

\def\x{\textbf{x}}
\def\d{\textrm{d}}
\def\n{\nabla}

\def\nn{\textbf{n}}
\def\i{\textrm{i}}

\def\ba#1\ea{\begin{align}#1\end{align}}

\def\nhat{\hat{\textbf{n}}}
\def\q{\textbf{q}}
\def\R{\mathcal{R}}
\def\i{\mathrm{i}}

\def\h{\hbar}

\def\v{\textbf{v}}

\def\bsa#1\esa{\begin{subequations}
\begin{align}#1\end{align} \end{subequations}}

\usepackage{graphicx}% Include figure files
\usepackage{dcolumn}% Align table columns on decimal point
\usepackage{bm}% bold math
\definecolor{darkblue}{RGB}{83,0,93}

\def\R{\mathbb{R}}
\begin{document}

\preprint{APS/123-QED}

% \title{}% Force line breaks with \\
%\title{Perfect and broadband cylindrical cloak for flexural waves}% Force line breaks with \\
%\thanks{myself}%

\title{Euler-Schr\"odinger Transformation}% Force line breaks with \\
%\title{Perfect and broadband cylindrical cloak for flexural waves}% Force line breaks with \\
%\thanks{myself}%

\author{Ahmad Zareei}
\affiliation{School of Engineering and Applied Sciences, Harvard University, Cambridge, MA, 02148}%\altaffiliation[]{SEAS, Pierce Hall}%Lines break automatically 
%  \altaffiliation[Also at]{}%Lines break automatically or can be forced with \\
%  \author{M.-Reza Alam}%
%  \email{alam@berkeley.edu}
%  \affiliation{%
%  Department of Mechanical Engineering, University of California, Berkeley\\

%\affiliation{
%\affiliation{
%\affiliation{%


%\date{}% It is always \today, today,

%%% Local Variables:
%%% mode: latex
%%% TeX-master: t
%%% End:

\begin{abstract}

%%%%%%%%%%%%%%%%%%%%%%% NEW Abstract
Here we present a transformation that maps the Schr\"odinger equation of quantum mechanics to the incompressible Euler equations of fluid mechanics. The transformation provides a wave solution and a potential function based on fluid properties that satisfy the Schr\"odinger equation given that the fluid velocity and pressure satisfy the Euler equations. Interestingly, in our transformation, the equivalent of quantum potential becomes the physical surface tension. This is contrary to the Madelung transformation that maps the Schr\"odinger equation to the compressible Euler equations where there is no physical counterpart for the quantum potential. Lastly, we show that using this transformation, the Bohm equation can be mapped to a particle's equation of motion moving on the free surface of the fluid.

% Here we present a transformation that maps the Sh\"odinger equation of quantum mechanics to  the incompressible Euler's equations of fluid mechanics with a free surface. The transformation provides a wave solution and a potential function based on fluid properties that satisfy the Schr\"odinger equation provided that the fluid velocity and pressure satisfy Euler's equations. Interestingly, in our transformation, the equivalent of quantum potential becomes the physical surface tension. This is contrary to Madelung transformation that maps the Schr\"odinger equations to compressible Euler's equation where there is no physical counterpart for the quantum potential. Lastly, we show that using this transformation, the Bohm equation can be mapped to a particle's equation of motion moving on the free surface of the fluid. 

% The transformation introduced here connects to the   the bouncing droplet experiments and show why this setup can be seen as  an analogue of quantum mechanics.

% \begin{description}

%\end{description}
\end{abstract}

% \pacs{Valid PACS appear here}% PACS, the Physics and Astronomy
\maketitle
%\tableofcontents
%%%%%%%%%%%%%%%%%%%%%%%%%%%%%%%%%%%%%%%%%%%%%%%%%%%%%%%%%%%%%%%%%%%%%%%%%%%%%%
%\section{Introduction}
%\label{intr}

We present a transformation between the Schr\"odinger
equation of quantum mechanics and the incompressible Euler equations (\textbf{incompressible}, inviscid, and irrotational hydrodynamic equations for the surface gravity waves). It is to be noted that the transformation between the Schr\"odinger equation and \textbf{compressible} Euler equations is known as the Madelung transform. Contrary to the Madelung transformation where there is no classical analog for the quantum potential, we find a classical counterpart to the quantum potential which becomes the surface tension with the Planck constant acting similar to the surface tension over density. We present two versions of our transformation, where we add complexity to the transformation step by step: (i) in the first version, we map the Schr\"odinger equation to the surface gravity wave equations for an inviscid, irrotational, and incompressible fluid with surface tension in deep water; (ii) next, we assume a finite depth fluid and also a particle bouncing on the free surface (bouncind droplet setup \cite[][]{couder2005dynamical}), and show how the transformation changes in mapping Schr\"odinger equation to the Euler equations. Assuming Bohmian mechanics, we find the analog of Bohm equation using our transformation, and show that under certain conditions, the transformation of Bohm equation matches with the Newton equation of motion for the bouncing droplet. This process may help us understand why the bouncing droplet problem~\cite[][]{couder2005dynamical} has shown quantum resemblance (such as single-slit patterns and double-slit diffraction~\cite[][]{couder2006single}, tunnelling~\cite[][]{eddi2009unpredictable}, orbital quantization~\cite[][]{fort2010path}, level splitting~\cite[][]{eddi2012level}, and wavelike statistics in a confined region~\cite[][]{harris2013wavelike}).

\section{(I) Simplified Euler-Schr\"odinger Transformation}
\label{sec:overview}

\textit{Schr\"odinger equation}-- The position-space Schr\"odinger equation for a single non-relativistic particle with mass $m$ in a potential $V(\q,t)$ is 
\begin{align} \label{A200}
  \mathrm{i} \hbar \f{\p \psi}{\p t} (\q,t) = -
  \f{\h^2}{2m} \Delta \psi(\q,t) + V(\q) \psi (\q,t)
\end{align}
 where $\hbar$ is the reduced Planck constant ($\hbar = h /2\pi$), and $\psi(\q,t): \R^3 \times \R \RA \mathbb{C}$  is a wave function that assigns a complex number to each point $\q \in \R^3$ in space at each time $t\in \R$.

\textit{Euler equations}-- Considering the irrotational motion of a homogeneous,
incompressible, and inviscid fluid with the surface tension
$\sigma$ and density $\rho$; the governing equations in terms of the velocity
potential $\phi$ and surface elevation $\eta$ read
\begin{subequations}\label{A410}
\begin{align} & \n^2 \phi = 0, \qquad z \leq \eta, \label{A410a} \\ &
\eta_t + \n (\eta -z). \n \phi = 0, \qquad z= \eta,\label{A410b} \\
&\phi_t + \f{1}{2} | \n \phi|^2 + g \eta + \f{\sigma}{\rho} \n. \nhat  = 0  \qquad z = \eta, \label{A410c} \\ & \phi_z = 0, \qquad z \RA
-\infty, \label{A410d}
\end{align}
\end{subequations}
where $\nhat$ is the unit normal vector at the surface $z=\eta$. In the
governing equations, \eqref{A410a} is the continuity
equation in the fluids domain, \eqref{A410b} and \eqref{A410d}
are kinematic boundary conditions at the surface and bottom, and
\eqref{A410c} is the dynamic boundary condition at the
surface. 

\begin{figure}[!h]
  \centering
  \includegraphics[width=0.45\textwidth]{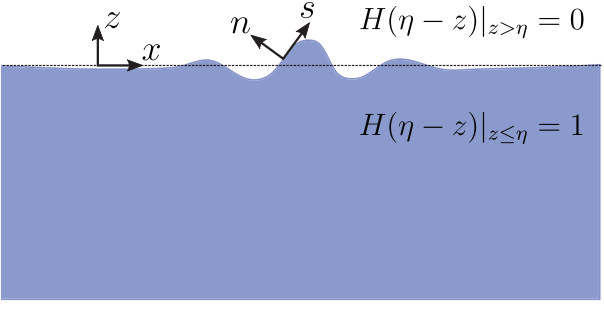}
  \caption{Schematic of an incompressible fluid with a free surface. The calm fluid surface is at $z=0$ and positive $z$ points upward. The tangential and normal coordinate system is defined on the free surface, where positive $n$ points outward. The step functions are unity inside the fluids domain
    $\mathcal{H} = H(\eta - z) |_{z<\eta}=1$, and zero outsize $H(\eta - z) |_{z>\eta}=0$.}\label{fig:100A}
\end{figure}

\textit{The Euler-Schr\"odinger Transformation}-- In the Schr\"odinger equation, we pick $m=1$, $\hbar = \sqrt{{2\sigma}/{\rho}}$ and define the potential function as
$V = \f{1}{\hbar} \lp P/\rho + gz\rp$,
where the parameters are related to the fluid equations of motion. We
assume a wave function in the form of
$\Psi = H(\eta - z) \exp\lp {\i \phi/\hbar}\rp$, where $H(\cdot)$ is a step
function (see Fig. \ref{fig:100A}).
Next, we show that if $\Psi= H(\eta - z) \exp\lp {\i \phi/\hbar}\rp$ satisfies the Schr\"odinger's equation (Eq.\eqref{A200}) with $V = \f{1}{\hbar} \lp P/\rho + gz\rp$, the fluid properties $\eta, \phi, \rho, P$ should then satisfy the Euler equations
(Eq. \eqref{A410}). Inserting the wave function into the Schr\"odinger
equation \eqref{A200}, we obtain the following equations for the
real and imaginary part:
\begin{subequations}\label{A50000ABC}
  \begin{align}
    & -H(\eta-z) \phi_t + \f{\sigma}{\rho} \n^2H (\eta -z) -\f{1}{2}
      H(\eta - z) |\n \phi|^2 \nonumber \\
      & \qquad - \lp\f{P}{\rho} + g z \rp \cdot H(\eta -z) =
      0, \label{A500ABCb}\\
        & \i \lp \f{\p }{\p t} H(\eta-z) + \n H(\eta-z)\cdot  \n
      \phi\rp + \i\f{1}{2} H(\eta -z) \n^2 \phi = 0. \label{A500ABCa} 
\end{align}
\end{subequations}
Here Eq. \eqref{A500ABCb} is the real part and Eq. \eqref{A500ABCa} is the imaginary part. Note that $F(x,y,z) := \eta(x,y) - z= 0$ defines the
free water surface $\Gamma_s$, and as a result
$\p H(\eta -z) / \p t = \eta_t \delta (\eta-z)$, and
$\n H(\eta-z) = \n(\eta-z) \delta(\eta - z)$. We now show that Eqs. \eqref{A50000ABC}a-d is equivalent to
Eq. \eqref{A410}. Multiplying Eq. \eqref{A500ABCa} by an
arbitrary function $f(\x)$ where $\x \in \mathbb{R}^{3}$ and
integrating over the whole space $\Omega\subseteq \mathbb{R}^{3}$, we
obtain
\begin{align}  
  & \iiint_\Omega \lb f(\mathbf{x}) \lp \eta_t + \n (\eta-z). \n \phi\rp \rb \delta(\eta -z)\d V \nonumber \\
  & \qquad + \iiint_{\Omega} \f{1}{2} H(\eta -z) f(\mathbf{x}) \n^2 \phi  \d V = 0,
\end{align}
which reduces to
\begin{align}
  & \iint_{\mathcal{S}} \lb f(\mathbf{x}) \lp \eta_t + \n (\eta-z). \n \phi \rp\rb _{z=\eta} \d A  \nonumber \\
  & \qquad + \iint_{\mathcal{S}} \d A \int_{-\infty}^{\eta}\d z f(\mathbf{x})\n^2\phi = 0,
\end{align}
where $\mathcal{S} \subseteq \mathbb{R}^2$ is the free surface of the
fluid. Since $f(\mathbf{x})$ is an arbitrary function, we
obtain
\begin{subequations} \label{A10-11}
\begin{align}
\n^2 \phi = 0 , \qquad -\infty \leq z \leq \eta \label{A1000}\\
\eta_t + \n (\eta-z). \n \phi=0 \qquad z =\eta \label{A1100}
\end{align}
\end{subequations}
which are respectively the incompressibility and the kinematic
boundary conditions. We now show that \eqref{A500ABCb} simplifies to the dynamic boundary
condition \eqref{A410c}. Let $g(\mathbf{x})$ be a function with
supp$(g) = \mathbb{R}^{2}\times [a,b]$, where
$a<\inf(\eta)-\delta <0$, $b>\sup(\eta)+\delta$, and $\delta>0$.  We
assume the normal derivative of this function on the surface is zero,
i.e. $\p g/\p n |_{z=\eta} = 0$, where $\nn$ is the normal direction
on the surface $z=\eta$.  We now multiply equation \eqref{A500ABCb} by
this function $g(\mathbf{x})$ and take the integral over
$\Omega = \mathbb{R}^{2}\times [a',b']$, where in the $z$-direction it
contains supp$(g)$, i.e.  $a'\leq a\leq 0\leq b\leq b'$. We find
\begin{align} 
  & \iiint_\Omega  H(\eta-z) g(\mathbf{x}) \lp \phi_t + \f{1}{2} |\n
  \phi|^2 + \frac{P}{\rho}  + g \rp \d V \nonumber \\
  & \qquad - \f{\sigma}{\rho}  \iiint_{\Omega} g(\mathbf{x})\n\
    \cdot\lp -\nn \delta (n) \rp  = 0, 
\end{align}
where we used the fact that $\n H(\eta-z) = -\nhat~\delta(n)$. Expanding the right hand side,  we obtain
\begin{align}
&     \iint_\mathcal{S} \d A \int_{a'}^{\eta} \d z g(\mathbf{x}) \lp \phi_t +
  \f{1}{2} |\n \phi|^2 + \frac{P}{\rho} + g \rp \nonumber \\
  & + \f{\sigma}{\rho} \lb  \iiint_{\Omega} \lp g(\mathbf{x}) \n\cdot\nhat \delta(n) + g(\mathbf{x})\f{\p}{\p n}\delta(n) \rp \d V \rb  = 0.
\end{align}
and as a result
\begin{align} \label{A1500}
  & \iint_\mathcal{S} \d A \int_{a'}^{\eta} \d z g(\mathbf{x}) \lp \phi_t -  \f{1}{2} |\n \phi|^2 + \frac{P}{\rho}  + g \rp \nonumber\\
& \quad  + \f{\sigma}{\rho}\iiint_{\Omega}  g(\mathbf{x}) \n\cdot\nhat \delta(n)\,\d V = - \f{\sigma}{\rho}    \iiint_{\Omega} g(\mathbf{x})\f{\p}{\p n} \delta(n) \, \d V.
\end{align}
The right-hand side of the above equation is zero: if we
use integral by parts on the right-hand side, we obtain 
\begin{align}
  & \iiint_{\Omega} g(\mathbf{x})\f{\p}{\p n} \delta(n)\d V \nonumber \\
  & \qquad = \iiint_{\Omega} \f{\p}{\p n} \lp g(\mathbf{x}) \delta(n) \rp \d V - 
\iiint_{\Omega}\lp \f{\p g(\mathbf{x})}{\p n}\rp \delta(n) \,\d V,
 % \\ & = \lim _{\mathcal{L}\rightarrow \infty} \iint_{\mathcal{R}} \d t \d s ~ g(\mathbf{x})\delta(n)|_{-\mathcal{L}}^{\mathcal{L}}
\end{align}
where (i) the first and (ii) the second integrals are zero, since (i)
by changing the coordinates from Cartesian coordinate $(x,y,z)$ to
$(t,s,n)$, this integral vanishes with the compactness of
$g(\mathbf{x})$ and (ii) we assumed that   $\p g/\p n|_{z=\eta}
=0$. Looking back at Eq. \eqref{A1500}, since $g(\mathbf{x})$ is arbitrary, we obtain
%
%\begin{subequations}
\begin{align}\label{A16-17}
%   &\phi_t - \frac{1}{2} |\n \phi|^2 + \frac{P}{\rho} + (g-\g) z =0, \quad z\leq \eta \\
  &\phi_t - \frac{1}{2} |\n \phi|^2 + \f{P}{\rho} + g z  +  \lp \frac{\sigma}{\rho} \n.\nn \rp \delta(n) = 0, \quad z \leq  \eta 
\end{align}
%\end{subequations}
%
which is the same as equation \eqref{A410c} on the surface. Note that  Eq. \eqref{A16-17} is what we obtain starting from Navier-Stokes equations, and in the Euler equations we only consider the above equation on the free surface, i.e., $z=\eta$. Equations \eqref{A10-11} and \eqref{A16-17} together are collectively the Euler equations (Eq. \eqref{A410}).  Note that, since the domain is unbounded in the
$z$-direction, we impose the regularizing condition that $\phi$ vanishing at infinity (i.e., $z\RA \infty$), which is Eq. \eqref{A410d} and is a condition for the velocity potential to be bounded.

%%%%%%%%%%%%%%%%%%%%%%%%%%%%%%%%%%%%%%%%%%%%%%%%%%%%%%%%%%%%%%%%%%%
%%%%%%%%%%%%%%%%%%%%%%%%%%%%%%%%%%%%%%%%%%%%%%%%%%%%%%%%%%%%%%%%%%%
%%%%%%%%%%%%%%%%%%%%%%%%%%%%%%%%%%%%%%%%%%%%%%%%%%%%%%%%%%%%%%%%%%%
%%%%%%%%%%%%%%%%%%%%%%%%%%%%%%%%%%%%%%%%%%%%%%%%%%%%%%%%%%%%%%%%%%%
%%%%%%%%%%%%%%%%%%%%%%%%%%%%%%%%%%%%%%%%%%%%%%%%%%%%%%%%%%%%%%%%%%%

\section{(II) Euler-Schr\"odinger transformation}

\textit{Schr\"odinger equation}-- As discussed in the previous section, the Schr\"odinger equation is 
\begin{align} \label{Schrodinger-200}
  \mathrm{i} \hbar \f{\p \psi}{\p t} (\x,t) = -
  \f{\h^2}{2m} \Delta \psi(\x,t) + V(\x) \psi (\x,t),
\end{align}
where $\hbar$ is the reduced Plank constant, and
$\psi(\x,t): \R^3 \times \R \RA \mathbb{C}$ is the wave function of a
 particle with mass $m$ in the  potential $V(\x)$. Considering Bohmian mechanics, there exists a particle with mass $m$ and position $\Q\in \R^3 $, and the particle is guided by a the wave function $\psi(\x,t)$ with the
equation
\begin{align}\label{Bohm-Particle100}
  \f{\d \Q}{\d t} = \f{\hbar}{m}  \Im \left\{ \f{\n \psi}{\psi} (\x ,t)\right\},
\end{align}
where $\Im(z)$ gives the imaginary part of the complex number
$z$. This equation is known as Bohm equation, and governs the
trajectory of a quantum particle.

\begin{figure}[!h]
  \centering
   \includegraphics[width=0.4\textwidth]{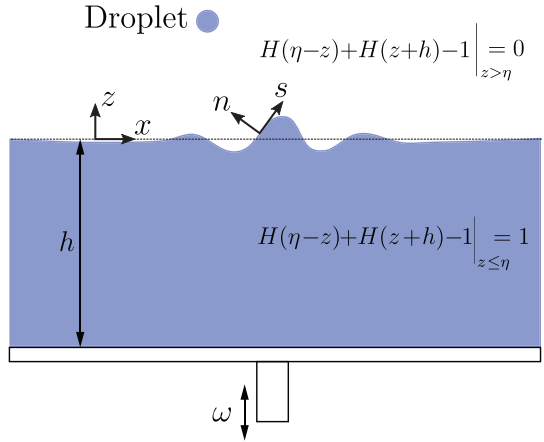}
  \caption{Schematic of the fluid bath and droplet bouncing on
    its free surface. The calm fluid surface is at $z=0$ with positive $z$ pointing upward. The tangential coordinate is defined on the surface with
    positive normal direction $n$ points out of the fluid's
    domain. The step function
    $\lb H(\eta - z) + H(z+h) -1\rb |_{-h<z<\eta} = 1$ inside the
    fluid and $\lb H(\eta - z) + H(z+h) -1\rb |_{z>\eta} = 0$ outside
    of the fluid domain. The fluid depth is
    $\tilde h$.}\label{fig:schematic-setup}
\end{figure}

\textit{Euler's equations}-- Now consider a fluid bath with a free surface on a vibrating plate where a droplet is bouncing on its free surface (see Fig. \ref{fig:schematic-setup}).  We assume a coordinate system attached to the vibrating bath with a positive $z$-axis pointing upward and the fluid's calm surface being at $z=0$. The vibration frequency of the
bath is identified by $\omega$ and the acceleration amplitude is
$\ggg$ i.e. the acceleration is 
$\g = \ggg \cos(\omega t)$.
%position is obtained as $(-\g/\omega^{2})\cos(\omega t)$.
We consider the fluid to be homogeneous, incompressible, 
inviscid, and irrotational. At each
time the droplet hits the surface, it receives some
force $\vec{\mathcal{F}}_i$ at the particle's location $\X_i$ and
impacts time $t_i$. Since $\vec{\mathcal{F}}_{i}$ is the contact force
applied to the droplet, $-\vec{\mathcal{F}}_{i}$ is the reaction of
this force and is applied to the fluid surface. The governing equations for the fluid read as
\begin{subequations}\label{410}
  \begin{align} 
    & \n^2 \phi = 0, \qquad z \leq \eta, \label{410a} \\
    & \eta_t + \n (\eta -z). \n \phi = 0, \qquad z= \eta,\label{410b} \\
    &\phi_t + \f{1}{2} | \n \phi|^2 + (g-\g) \eta + \f{\sigma}{\rho} \n. \nhat + \nonumber\\
    & \qquad \sum_{i} \Theta_{i} \delta(\x-\X_i)
      \delta(t-t_i) = 0 \qquad z = \eta, \label{410c} \\
    & \phi_z = 0, \qquad z = - h, \label{410d}
\end{align}
\end{subequations}
where $z=-h$ is the bottom of the fluid bath, $\nhat$ is unit normal
vector at the surface $z=\eta$, and $\Theta_i$ is a potential function
defined as $\vec{\mathcal{F}}_{i} = - \n \Theta_i$. Here, \eqref{410a}
is the continuity equation in the fluid domain, \eqref{410b} and
\eqref{410d} are kinematic boundary conditions at the free surface and
bottom boundary, and \eqref{410c} is the dynamic boundary condition at
the surface. The Euler equations (Eqs. \eqref{410}a-d) and droplet's equation
are coupled, and should be solved together. The droplet's equation of
motion can be written as
\begin{align}\label{AAA300}
  m \frac{\d^{2} \X}{\d t^{2}} = -m(g - \g)\, \hat{\mathbf{z}}+ \sum_{i} \vec{\mathcal{F}}_i \delta(\x - \X_i)\delta(t-t_{i}),
\end{align}
where $m$ is the mass of the droplet, $\X$ is its position vector of
the droplet,
$\vec{\mathcal{F}}_i$ accounts for the contact force applied to the
particle at time $t_i$ and the particle's location $\X_i$, and
$\delta(\cdot)$ is the Dirac delta function. It is worth noting that, in the Eqs. \eqref{410}a-d, if we set
all $\vec{\mathcal{F}}_i=0$, and also $\g = 0$, the governing
equations reduce to general Euler equations (Eqs. \eqref{A410}a-d).

\textit{Euler-Schr\"odinger Transformation}-- In this section, we present a wave function $\Psi$ and a potential function $V$ as a function of fluid free surface $\eta$, and velocity potential $\phi$, such that if $\Psi$ and $V$ satisfy the Schr\"odinger equation (Eq. \eqref{Schrodinger-200}) then the resulting $\eta$ and $\phi$ satisfy the Euler equations (Eqs. 
\eqref{410}a-d). Conversely, if $\eta$ and $\phi$ satisfy the Euler
equations, the obtained wave function $\Psi$ satisfies the
Schr\"odinger equation with the potential function $V$.
% We would like to point out that in the previous section, the surface
% wave equation is coupled with the droplet's motion, our argument
% here is also true, even if we only solve for the surface waves
% i.e. assuming $\mathcal{F}= 0 = \Theta $.
%

Consider a wave function as
\begin{equation}
  \label{eq:1}
  \Psi \equiv \lb H(\eta - z) + H(z+h) -1\rb \exp (\i \phi/\hbar),
\end{equation}
where $H(\cdot)$ is a step function, and $z=-h$ represents the bottom
of the fluid bath. Additionally, we define a potential function $V$  in the
Schr\"odinger equation based on the fluids parameters as 
\begin{align}
  V \equiv \f{1}{\hbar} \lp P/\rho + (g-\g)z + \sum \Theta_i \delta(\x-\X_i)
\delta(t-t_i) \rp,
\end{align}
and assume $m\equiv 1$, $\hbar \equiv \sqrt{2\sigma/\rho}$. Inserting
the wave function into the Schr\"odinger equation, we obtain
\begin{align} \label{A150000}
  & \i \lp \f{\p }{\p t} H(\eta-z) + \n H(\eta-z). \n
    \phi\rp +  \nonumber \\ & \quad \i \lp \f{\p }{\p t} H(z+h) + \n H(z+h). \n
    \phi\rp + \nonumber\\
 & \qquad \quad \i\f{1}{2} \lb H(\eta -z) + H(z+h) -1 \rb \n^2 \phi = 0,
\end{align}
for the imaginary part, and for the real part we obtain
\begin{align} \label{A250000}
  & - \lb H(\eta-z) + H(z+h) -1 \rb \phi_t + \f{\sigma}{\rho} \n^2H (\eta -z) \nonumber \\ & \quad + \f{\sigma}{\rho} \n^2H (z+h)  - \lb H(\eta -z) + H(z+h) -1
    \rb \cdot \nonumber \\
  & \lp \frac{1}{2}|\n \phi|^2 + \f{P}{\rho} + (g-\g) z + \sum
    \Theta_{i}\delta(\x-\X_i)\delta(t-t_i)\rp = 0.
\end{align}
Note that Eqs. \eqref{A150000} and \eqref{A250000} here are similar to the
Eqs. \eqref{A500ABCa} and \eqref{A500ABCb}. 
% Additionally,  in  Eq. \eqref{A250000}, the term $\n^2H(z+h) = 0$ since the bottom topography is flat.
Equation \eqref{A150000} can be simplified to
\begin{align}
    & \i \lp \frac{\p \eta}{\p t}  + \n (\eta-z). \n
    \phi\rp \delta (\eta - z)  +  \i \lp \hat{\mathbf{z}}. \n \phi \rp\delta (z+h)  + \nonumber\\
 & \qquad \quad \i\f{1}{2} \lb H(\eta -z) + H(z+h) -1 \rb \n^2 \phi = 0.
\end{align}
With an argument, similar to that of the previous section, this
equation reduces to
\begin{subequations}
\begin{align}
  & \n^2\phi = 0, \quad -h \leq z \leq \eta ,\label{similar-to-410a}\\
  &\eta_t + \n (\eta - z). \n \phi = 0, \quad z= \eta, \label{similar-to-410b}\\
  & \phi_z = 0, \qquad z = -h, \label{similar-to-410d}
\end{align}
\end{subequations}
where equations \eqref{similar-to-410a}, \eqref{similar-to-410b},
\eqref{similar-to-410d} matches with equations \eqref{410a},
\eqref{410b}, \eqref{410d} respectively. Furthermore, equation
\eqref{A250000}, with an argument similar to that of the previous
section, reduces to
\begin{subequations}
\begin{align}
  & \phi_t + \frac{1}{2}|\n \phi|^2  + (g-\g) z + \nonumber \\ & \qquad  \sum
    \Theta_{i}\delta(\x-\X_i)\delta(t-t_i)   + \lp  \frac{\sigma}{\rho} \n. \nn \rp   = 0, \quad  z = \eta, \\
  & \phi_t + \frac{1}{2}|\n \phi|^2 + \f{P}{\rho} + (g-\g) z + \nonumber \\ & \quad  \sum
    \Theta_{i}\delta(\x-\X_i)\delta(t-t_i)    = 0, \quad -h \leq z < \eta,
\end{align}
\end{subequations}
which at $z =\eta$ matches with equation \eqref{410c}. Note that the fluid
pressure becomes zero at
the fluid's free surface, i.e., $P_{z=\eta}=0$.

\textit{Particle's Motion}-- Inserting the wave function \eqref{eq:1} into the Bohm equation
\eqref{Bohm-Particle100}, we obtain
\begin{align}\label{eq:velocity-potential-Bohm}
\v = \f{\d \Q}{\d t} = \n \phi.
\end{align}
Note that in our droplet's setup, the droplet is outside the fluid,
while $\phi$ is only defined at the surface and inside the fluid
domain. We, therefore, take $\phi$ as the velocity potential
calculated at the surface. Taking the time derivative of equation
\eqref{eq:velocity-potential-Bohm}, the particle's dynamic equation is obtained. Using 
\eqref{410b}, the acceleration of the particle becomes
\begin{align}
  \f{\d \v}{\d t} & = \n \phi_t  =  \n \lp \f{1}{2}|\n \phi|^2\rp  - (g-\g)\, \hat{\mathbf{z}} +  \nonumber\\
                  &   \sum \vec{\mathcal{F}}_i \delta(\x-\X_i) \delta(t-t_i) +  \n \lp \frac{\sigma}{\rho} \n.\nn \rp. 
\end{align}
Assuming the non-linearity is small i.e.  $|\n \phi|/|\v|\ll 1$, the
droplet's equation of motion becomes
\begin{align}
  \f{\d \v}{\d t} =  - (g-\g)\, \hat{\mathbf{z}} + \sum \vec{\mathcal{F}} \delta(\x-\X_i) \delta(t-t_i)+  \n \lp \frac{\sigma}{\rho} \n.\nn \rp.
\end{align}
The obtained equation is similar to \eqref{AAA300} with an extra term
on the right-hand side, that is related to the gradient effect of surface
tension. 

%%%%%%%%%%%%%%%%%%%%%%%%%%%%%%%%%%%%%%%%%%%%%%%%%%%%%%%%%%

\section{Remarks}

In this section, we review some basic conclusions that can be made
based on the transformation.

\begin{itemize}
\item The probability of finding a particle at a point is proportional
  to $|\psi|^2$ and the integral of this probability over the whole
  space should add up to unity 
  \begin{align}
  \int |\psi|^2\d V  = 1.
  \end{align} 
  The equivalent of this condition, using the transformation,
  translates into the fact that the volume of the fluid is constant,
  \begin{align}
  \int H(\eta - z)\cdot H(\eta - z) \d V = \int_{\Lambda} \d V = V_0,
  \end{align}
  where $\Lambda$ is the fluids domain and $V_0$ is the
  total volume of fluid.

\item The flow of probability using the Schr\"odinger equation is obtained
  as
  \begin{align}
    \frac{\p |\psi|^2}{\p t} + \n . \lb \frac{\hbar}{2m\i} \lp \psi^{*}\n \psi - \psi \n \psi^{*} \rp \rb = 0.
  \end{align}
  The equivalent form of this equation, using the Euler-Schr\"odinger equation
  becomes
  \begin{align}
    &\qquad  \frac{\p H(\eta - z)\cdot H(\eta - z)}{\p t}+ \nonumber \\ &\qquad \quad \quad \frac{1}{2} \n. (H(\eta - z)\cdot H(\eta - z) \n \phi) = 0, \\
    & \qquad \delta (\eta - z) \lb \eta_t + \n (\eta - z) . \n \phi \rb + \nonumber \\ & \qquad \quad \quad \frac{1}{2} H(\eta - z) \n^2\phi = 0,
  \end{align}
  which is the same as surface kinematic boundary condition together
  with the Laplacian equation (conservation of mass).

\item The scale at which quantum effects are observed is such that
  $\hbar/ML^2T^{-1}\approx 1$, where $M,L$, and $T$ are the mass, length
  and time scales of the problem. Similar arguments for the bouncing
  droplet's problem, yield $\sigma/\rho g L^2\approx 1$. As a
  result, the length scale at which we expect to see quantum-like
  behaviors for the silicone oil becomes $L\approx 2.5 mm$, which
  is close to  the length scale of experiments.
\end{itemize}

\section{Conclusion}
In summary, we provided a transformation that maps the
Schr\"odinger equation of quantum wave mechanics to Euler equations
of fluid wave mechanics. Specifical,ly we showed
\begin{widetext}
\begin{align}
  \text{Schr\"odinger Equation:} & \qquad \mathrm{i} \hbar \f{\p \psi}{\p t}(\x,t)  = -
  \f{\h^2}{2m} \Delta \psi(\x,t)  + V(\x)  \psi (\x,t) \label{eq:schrodinger-conclustion} \\
  \text{Assuming:} \quad \Bigg\updownarrow &  \psi(\x,t) = \lb H(\eta - z) + H(z+h) -1 \rb e^{\i \phi/ \hbar}, ~ V = \frac{1}{\hbar} \lp \frac{P}{\rho} + g z \rp \label{eq:transformation-conclusio} \\
  \text{Euler's Equation:} & \qquad 
  \begin{cases}
    &  \n^2 \phi =0, \qquad -h \leq z \leq \eta\\
    & \eta_t + \n (\eta -z). \n \phi = 0, \qquad z= \eta \\
    & \phi_t + \f{1}{2} | \n \phi|^2 + g  \eta + \f{\sigma}{\rho} \n. \nhat  = 0  \\
    & \phi_z = 0, \qquad z = - h 
  \end{cases} \label{eq:euler-equations-conclusion}
\end{align}
\end{widetext}
wherein \eqref{eq:schrodinger-conclustion}, $\psi$ is the wave
function, $\hbar$ is the reduced Plank constant, and $m$ is the
quantum particle's mass; and in \eqref{eq:euler-equations-conclusion},
$\phi$ is the  velocity potential, $z=\eta $ is the free surface, $\rho$ is fluid's density, $\sigma$ is the surface tension, and $\nn$ is the
a normal vector pointing outward of the fluid's free surface at
$z= \eta$. Equation \eqref{eq:transformation-conclusio} also provides
the transformation that relates the quantum wave solution and
potential to the fluid's parameters. We further showed that the same transformation works even when the
fluid bath is vibrating with the acceleration
$\gamma =\hat{\gamma} \cos \omega t$ and there is a particle that
bounces on the surface and exerts a force at the
surface $\mathcal{F}_i = \n \Theta_i$ at the point $\x = \X_i$ and time
$t=t_i$. Precisely the equations are
\begin{widetext}
\begin{align}
  \text{Schr\"odinger Equation:} & \qquad \mathrm{i} \hbar \f{\p \psi}{\p t}(\x,t)  = -
  \f{\h^2}{2m} \Delta \psi(\x,t)  + V(\x)  \psi (\x,t) \nonumber \\
  \text{Assuming:} \quad \Bigg\uparrow &  \psi(\x,t) = \lb H(\eta - z) + H(z+h) -1 \rb e^{\i \phi/ \hbar},\nonumber \\ 
 \text{and} \quad \quad \Bigg\downarrow  & ~ V = \frac{1}{\hbar} \lp \frac{P}{\rho} + (g-\g)z + \sum \Theta_i\delta(\x- \X_i) \delta (t-t_i)\rp \nonumber \\
  \text{Euler's Equation:} & \qquad 
  \begin{cases}
    &  \n^2 \phi =0, \qquad -h \leq z \leq \eta\\
    & \eta_t + \n (\eta -z). \n \phi = 0, \qquad z= \eta \\
    & \phi_t + \f{1}{2} | \n \phi|^2 + (g-\g) \eta + \f{\sigma}{\rho} \n. \nhat +  \sum_{i} \Theta_{i} \delta(\x-\X_i) = 0  \\
    & \phi_z = 0, \qquad z = - h 
  \end{cases} \nonumber
\end{align}
\end{widetext}
Considering this transformation, the Bohm particle equation also maps to the particle's equation of motion if the surface remains almost flat and we can ignore the nonlinear term, i.e.,
\begin{align*}
  &\frac{\d \Q}{\d t}  = \frac{\hbar}{m} \Im \left\{ \frac{\n \psi}{\psi} (\x,t) \right\}\nonumber \\
  & \qquad \qquad \big \downarrow \nonumber \\
  & m \frac{\d^{2}\X}{\d t^2} = -m (g-\gamma) \hat{\mathbf{z}} + \sum \vec{\mathcal{F}}_i \delta(\x - \X_i) \delta (t- t_i) \nonumber
\end{align*}
%

% \bibliography{references.bib} 

\begin{thebibliography}{6}%
\makeatletter
\providecommand \@ifxundefined [1]{%
 \@ifx{#1\undefined}
}%
\providecommand \@ifnum [1]{%
 \ifnum #1\expandafter \@firstoftwo
 \else \expandafter \@secondoftwo
 \fi
}%
\providecommand \@ifx [1]{%
 \ifx #1\expandafter \@firstoftwo
 \else \expandafter \@secondoftwo
 \fi
}%
\providecommand \natexlab [1]{#1}%
\providecommand \enquote  [1]{``#1''}%
\providecommand \bibnamefont  [1]{#1}%
\providecommand \bibfnamefont [1]{#1}%
\providecommand \citenamefont [1]{#1}%
\providecommand \href@noop [0]{\@secondoftwo}%
\providecommand \href [0]{\begingroup \@sanitize@url \@href}%
\providecommand \@href[1]{\@@startlink{#1}\@@href}%
\providecommand \@@href[1]{\endgroup#1\@@endlink}%
\providecommand \@sanitize@url [0]{\catcode `\\12\catcode `\$12\catcode
  `\&12\catcode `\#12\catcode `\^12\catcode `\_12\catcode `\%12\relax}%
\providecommand \@@startlink[1]{}%
\providecommand \@@endlink[0]{}%
\providecommand \url  [0]{\begingroup\@sanitize@url \@url }%
\providecommand \@url [1]{\endgroup\@href {#1}{\urlprefix }}%
\providecommand \urlprefix  [0]{URL }%
\providecommand \Eprint [0]{\href }%
\providecommand \doibase [0]{http://dx.doi.org/}%
\providecommand \selectlanguage [0]{\@gobble}%
\providecommand \bibinfo  [0]{\@secondoftwo}%
\providecommand \bibfield  [0]{\@secondoftwo}%
\providecommand \translation [1]{[#1]}%
\providecommand \BibitemOpen [0]{}%
\providecommand \bibitemStop [0]{}%
\providecommand \bibitemNoStop [0]{.\EOS\space}%
\providecommand \EOS [0]{\spacefactor3000\relax}%
\providecommand \BibitemShut  [1]{\csname bibitem#1\endcsname}%
\let\auto@bib@innerbib\@empty
%</preamble>
\bibitem [{\citenamefont {Couder}\ \emph {et~al.}(2005)\citenamefont {Couder},
  \citenamefont {Protiere}, \citenamefont {Fort},\ and\ \citenamefont
  {Boudaoud}}]{couder2005dynamical}%
  \BibitemOpen
  \bibfield  {author} {\bibinfo {author} {\bibfnamefont {Y.}~\bibnamefont
  {Couder}}, \bibinfo {author} {\bibfnamefont {S.}~\bibnamefont {Protiere}},
  \bibinfo {author} {\bibfnamefont {E.}~\bibnamefont {Fort}}, \ and\ \bibinfo
  {author} {\bibfnamefont {A.}~\bibnamefont {Boudaoud}},\ }\href@noop {}
  {\bibfield  {journal} {\bibinfo  {journal} {Nature}\ }\textbf {\bibinfo
  {volume} {437}},\ \bibinfo {pages} {208} (\bibinfo {year}
  {2005})}\BibitemShut {NoStop}%
\bibitem [{\citenamefont {Couder}\ and\ \citenamefont
  {Fort}(2006)}]{couder2006single}%
  \BibitemOpen
  \bibfield  {author} {\bibinfo {author} {\bibfnamefont {Y.}~\bibnamefont
  {Couder}}\ and\ \bibinfo {author} {\bibfnamefont {E.}~\bibnamefont {Fort}},\
  }\href@noop {} {\bibfield  {journal} {\bibinfo  {journal} {Physical review
  letters}\ }\textbf {\bibinfo {volume} {97}},\ \bibinfo {pages} {154101}
  (\bibinfo {year} {2006})}\BibitemShut {NoStop}%
\bibitem [{\citenamefont {Eddi}\ \emph {et~al.}(2009)\citenamefont {Eddi},
  \citenamefont {Fort}, \citenamefont {Moisy},\ and\ \citenamefont
  {Couder}}]{eddi2009unpredictable}%
  \BibitemOpen
  \bibfield  {author} {\bibinfo {author} {\bibfnamefont {A.}~\bibnamefont
  {Eddi}}, \bibinfo {author} {\bibfnamefont {E.}~\bibnamefont {Fort}}, \bibinfo
  {author} {\bibfnamefont {F.}~\bibnamefont {Moisy}}, \ and\ \bibinfo {author}
  {\bibfnamefont {Y.}~\bibnamefont {Couder}},\ }\href@noop {} {\bibfield
  {journal} {\bibinfo  {journal} {Physical review letters}\ }\textbf {\bibinfo
  {volume} {102}},\ \bibinfo {pages} {240401} (\bibinfo {year}
  {2009})}\BibitemShut {NoStop}%
\bibitem [{\citenamefont {Fort}\ \emph {et~al.}(2010)\citenamefont {Fort},
  \citenamefont {Eddi}, \citenamefont {Boudaoud}, \citenamefont {Moukhtar},\
  and\ \citenamefont {Couder}}]{fort2010path}%
  \BibitemOpen
  \bibfield  {author} {\bibinfo {author} {\bibfnamefont {E.}~\bibnamefont
  {Fort}}, \bibinfo {author} {\bibfnamefont {A.}~\bibnamefont {Eddi}}, \bibinfo
  {author} {\bibfnamefont {A.}~\bibnamefont {Boudaoud}}, \bibinfo {author}
  {\bibfnamefont {J.}~\bibnamefont {Moukhtar}}, \ and\ \bibinfo {author}
  {\bibfnamefont {Y.}~\bibnamefont {Couder}},\ }\href@noop {} {\bibfield
  {journal} {\bibinfo  {journal} {Proceedings of the National Academy of
  Sciences}\ }\textbf {\bibinfo {volume} {107}},\ \bibinfo {pages} {17515}
  (\bibinfo {year} {2010})}\BibitemShut {NoStop}%
\bibitem [{\citenamefont {Eddi}\ \emph {et~al.}(2012)\citenamefont {Eddi},
  \citenamefont {Moukhtar}, \citenamefont {Perrard}, \citenamefont {Fort},\
  and\ \citenamefont {Couder}}]{eddi2012level}%
  \BibitemOpen
  \bibfield  {author} {\bibinfo {author} {\bibfnamefont {A.}~\bibnamefont
  {Eddi}}, \bibinfo {author} {\bibfnamefont {J.}~\bibnamefont {Moukhtar}},
  \bibinfo {author} {\bibfnamefont {S.}~\bibnamefont {Perrard}}, \bibinfo
  {author} {\bibfnamefont {E.}~\bibnamefont {Fort}}, \ and\ \bibinfo {author}
  {\bibfnamefont {Y.}~\bibnamefont {Couder}},\ }\href@noop {} {\bibfield
  {journal} {\bibinfo  {journal} {Physical review letters}\ }\textbf {\bibinfo
  {volume} {108}},\ \bibinfo {pages} {264503} (\bibinfo {year}
  {2012})}\BibitemShut {NoStop}%
\bibitem [{\citenamefont {Harris}\ \emph {et~al.}(2013)\citenamefont {Harris},
  \citenamefont {Moukhtar}, \citenamefont {Fort}, \citenamefont {Couder},\ and\
  \citenamefont {Bush}}]{harris2013wavelike}%
  \BibitemOpen
  \bibfield  {author} {\bibinfo {author} {\bibfnamefont {D.~M.}\ \bibnamefont
  {Harris}}, \bibinfo {author} {\bibfnamefont {J.}~\bibnamefont {Moukhtar}},
  \bibinfo {author} {\bibfnamefont {E.}~\bibnamefont {Fort}}, \bibinfo {author}
  {\bibfnamefont {Y.}~\bibnamefont {Couder}}, \ and\ \bibinfo {author}
  {\bibfnamefont {J.~W.}\ \bibnamefont {Bush}},\ }\href@noop {} {\bibfield
  {journal} {\bibinfo  {journal} {Physical Review E}\ }\textbf {\bibinfo
  {volume} {88}},\ \bibinfo {pages} {011001} (\bibinfo {year}
  {2013})}\BibitemShut {NoStop}%
\end{thebibliography}

%

\end{document}